\documentstyle[twoside,fleqn,espcrc2,epsf]{article}

\newcommand{\lw}[1]{\smash{\lower2.ex\hbox{#1}}}
\def\simlt{\rlap{\lower 3.5 pt\hbox{$\mathchar \sim$}}\raise 1pt \hbox {$<$}}
\def\simgt{\rlap{\lower 3.5 pt\hbox{$\mathchar \sim$}}\raise 1pt \hbox {$>$}}
\def\lats{Nucl. Phys. B (Proc. Suppl.) 53 (1997) }
\def\latf{Nucl. Phys. B (Proc. Suppl.) 47 (1996) }

\newcommand{\AmS}{{\protect\the\textfont2
  A\kern-.1667em\lower.5ex\hbox{M}\kern-.125emS}}

\title{%
\vspace{-2.8cm}
\begin{flushleft}
       {\normalsize UTCCP-P-31,\ UTHEP-374}   \\[-0.2cm]
       {\normalsize November 1997}   \\
\end{flushleft}
       \vspace{0.7cm}
Light Hadron Spectroscopy\thanks{Plenary talk
 presented at ``Lattice 97'', Edinburgh, Scotland, 22--26 July 1997.}}

\author{T.~Yoshi\'e\address{
Center for Computational Physics,
University of Tsukuba, Tsukuba, Ibaraki 305, Japan}}

\begin{document}

\begin{abstract}
Recent developments in calculations of the light hadron
spectrum are reviewed. 
Particular emphasis is placed on discussion of 
to what extent the quenched spectrum agrees with experiment.
Recent progress, both for quenched and full QCD, 
in reducing scaling violation with the use of improved
actions is presented.

\end{abstract}

\maketitle

\section{Introduction}

Deriving the light hadron spectrum from the first principles of
QCD has been a major subject of lattice QCD
simulations\cite{ref:reviews}.
A precise determination of the known hadron spectrum
would lead us to a fundamental verification of QCD.
We should also clarify the nature of observed hadrons,
provide predictions for hadrons not in the quark model,
and give informations for quantities of phenomenological
importance.

In order to achieve these goals, understanding and control
of various systematic errors are required.
One of major sources of systematic errors is that of
a finite lattice spacing.
Recent progress in reducing this systematic error
has been made in two ways.
For the quenched QCD spectrum,
development of computer power has enabled to push simulations 
toward smaller lattice spacings on physically larger lattices with 
higher statistics than the previous attempts.
As a result we are now in a status to discuss the problem of how well
quenched QCD describes the experimental spectrum.
Another progress in reducing scaling violation is brought 
with the use of improved quark actions.
Tests of improvement, previously made mainly in quenched QCD, 
have been extended this year to full QCD.

Finite size effects and chiral extrapolations have been
studied extensively in the past.
Several studies to investigate these systematic errors were 
also reported at the Symposium.

In this review we attempt to describe the present status of
spectroscopic studies.
Progress in quenched QCD spectrum is summarized in
sec.~\ref{sec:progress}, emphasizing results in the continuum limit.
Discussions on several issues in spectroscopic studies follow
in sec.~\ref{sec:issues},
which include study of finite size effects, chiral extrapolations,
and quenching error in meson decay constants.
After discussions on improvement of quark actions
in sec.~\ref{sec:improve},
attempts toward a realistic calculation in full QCD are presented
in sec.~\ref{sec:fullQCD}.
Sec.~\ref{sec:other} is devoted to results for masses of glueballs
and exotics.  Our conclusions are given in sec.~\ref{sec:conclusions}.

\section{Progress in Quenched QCD Spectrum}\label{sec:progress}
\subsection{major simulations}

Recent quenched simulations made with the plaquette gauge action
are compiled in Table \ref{tab:table-q}. 
See sec.~\ref{sec:improve} for those with improved 
gauge actions. 

Deriving precise quenched results in the continuum limit 
is a first step toward understanding the light hadron spectrum. 
The GF11 collaboration\cite{ref:GF11mass} carried out
the first systematic effort to achieve this goal 
with the Wilson quark action 
using three lattices with $a^{-1} = 1.4-2.8$ GeV  and 
the spatial size $La \approx 2.3$ fm.
 
This year the CP-PACS collaboration reported further effort 
in this direction\cite{ref:CPPACS}.
They made high statistics simulations
on four lattices with $a^{-1} = 2.0 - 4.2$ GeV and $La \approx 3$ fm.
Hadron masses are calculated for five quark masses
corresponding to $m_\pi/m_\rho$ = 0.75, 0.7, 0.6, 0.5
and 0.4, the last point being closer to the chiral limit
than ever attempted for the Wilson action. 
They reported continuum values of hadron masses with a statistical 
error of 0.5 \% for mesons and 1--3 \% for baryons.

Another trend in this year's simulations is a pursuit 
of reduction of scaling violation with the use of 
the Sheikholeslami-Wohlert\cite{ref:SW} or clover action.
Efforts in this direction were made by 
the UKQCD\cite{ref:UKQCDlat96,ref:UKQCDb57,ref:UKQCDlat97} 
and JLQCD \cite{ref:JLQCD-fB} collaborations
for the tadpole-improved\cite{ref:LM} clover action
and by the UKQCD, QCDSF\cite{ref:QCDSFlatest} and APETOV\cite{ref:APETOV}
collaborations for the non-perturbatively $O(a)$-improved\cite{ref:NPI} 
clover action (see also Ref.\cite{ref:Wittig} on this subject).
These studies have not yet reached the level of simulations with the Wilson 
action, being restricted to the parameter range 
$m_\pi/m_\rho \simgt 0.5$, $a^{-1} \simlt 3$ GeV, and
$La \simlt 2.0$ fm.

For the Kogut-Susskind (KS) quark action, 
the MILC collaboration\cite{ref:MILClat96} last year 
reported a result of nucleon mass in the continuum limit based on
simulations on four lattices with $a^{-1}=0.6 - 2.4$ GeV and 
$La \approx 2.7$ fm.
Not much progress has been made this 
year\cite{ref:MILC-Tsukuba-Gottlieb,ref:KimOhta}.

\begin{table*}[t]
\caption{Recent spectrum runs in quenched QCD
with the standard gauge action. 
New results since Lattice 96 are marked by double asterisks and
those with increased statistics by asterisks.
Quark actions are denoted in parentheses by
W: Wilson, C: clover, and KS: Kogut-Susskind. 
Clover coefficients are denoted by 1: tree level, TP: tadpole improved,
TP1: one-loop tadpole improved, and NP: non-perturbatively improved.}
\label{tab:table-q}
\begin{center}
\begin{tabular}{lrrrrrrr}
        &  $\beta$   & size &  (fm) & \#conf. & \ \ \ $m_\pi/m_\rho$   & \#m & ref.\\
\hline
\hline
MILC (W)** & 5.70 & $(12-24)^3\times48$ & 1.7--3.4 & 404-170 & 0.90-0.50 & 6  & 
\cite{ref:MILC-Tsukuba-Gottlieb,ref:MILClat97} \\
\hline
CP-PACS (W)**  & 5.90 & $32^3\times56$ & 3.21 & 800 & 0.75-0.40 & 5 & \cite{ref:CPPACS}\\
CP-PACS (W)**  & 6.10 & $40^3\times70$ & 3.04 & 600 & 0.75-0.40 & 5 & \cite{ref:CPPACS}\\
CP-PACS (W)**  & 6.25 & $48^3\times84$ & 3.03 & 420 & 0.75-0.40 & 5 & \cite{ref:CPPACS}\\
CP-PACS (W)**  & 6.47 & $64^3\times112$& 3.03 &  91 & 0.75-0.40 & 5 & \cite{ref:CPPACS}\\
\hline
\hline
UKQCD (C=TP)  & 5.70 & $(12,16)^3\times24$ & (2.1,2.8) & (482,145) & 0.78,0.65 & 2 & 
\cite{ref:UKQCDlat96,ref:UKQCDb57} \\
UKQCD (C=TP)  & 6.00 & $16^3\times48$ & 1.6  & 499 & 0.76-0.62 & 3 & 
\cite{ref:UKQCDlat96,ref:UKQCDlat97}\\
UKQCD (C=TP)* & 6.20 & $24^3\times48$ & 1.8  & 218 & 0.75-0.49 & 3 & 
\cite{ref:UKQCDlat96,ref:UKQCDlat97} \\
UKQCD (C=NP)** & 6.00 & $(16,32)^3\times48$ & (1.7,3.3)  & (497,70) & 0.77-0.50 & 3 & 
\cite{ref:UKQCDlat97}\\
UKQCD (C=NP)** & 6.20 & $24^3\times48$ & 1.7  & 251 & 0.71-0.54 & 3 & 
\cite{ref:UKQCDlat97} \\
\hline
QCDSF (C=1)** & 5.70& $16^3\times32$ & 2.4 & & 0.66-0.44 & 3 & 
\cite{ref:QCDSFlatest} \\
QCDSF (C=NP)** & 5.70& $16^3\times32$ & 3.3& & 0.77-0.56 & 6 & 
\cite{ref:QCDSFlatest} \\
QCDSF (W)*     & 6.00& $(16,24)^3\times32$ & (1.4,2.0) & O(5000,100)& 0.93-0.50 
& (4,3) & \cite{ref:QCDSFlatest,ref:QCDSFb60}\\
QCDSF (C=NP)* & 6.00& $(16,24)^3\times32$ & (1.7,2.6) & O(1000,200) & 0.90-0.41 & 
(6,3) & \cite{ref:QCDSFlatest,ref:QCDSFb60} \\
QCDSF (W)**    & 6.20& $24^3\times48$ & 1.6 & O(100) & 0.94-0.61 & 5 & 
\cite{ref:QCDSFlatest}\\
QCDSF (C=NP)** & 6.20& $24^3\times48$ & 1.8 & O(300) & 0.90-0.59  & 5 & 
\cite{ref:QCDSFlatest}\\
QCDSF (C=NP)** & 6.20& $32^3\times64$ & 2.4 & O(40) & 0.55-0.39  & 3 &
\cite{ref:QCDSFlat97-2} \\
\hline
APETOV (W)**& 6.20 & $24^3\times48$ & 1.7 & 50 & & 7 & 
\cite{ref:APETOV} \\
APETOV (C=NP)**&6.20 & $24^3\times48$ & 1.9 & 50 & 0.98-0.56 & 7 & 
\cite{ref:APETOV} \\
\hline
JLQCD (C=TP1)** & 5.90& $16^3\times40$ & 2.0 & 400 & 0.76-0.56 & 4 & 
\cite{ref:JLQCD-fB}\\
JLQCD (C=TP1)** & 6.10& $24^3\times64$ & 2.1 & 200 & 0.77-0.50 & 4 & 
\cite{ref:JLQCD-fB}\\
JLQCD (C=TP1)** & 6.30& $32^3\times80$ & 2.2 & 100 & 0.81-0.52 & 4 & 
\cite{ref:JLQCD-fB}\\
\hline
\hline
Kim-Ohta (KS)* & 6.50 & $48^3\times64$ & 2.6 & 350 & 0.65-0.28& 4 & 
\cite{ref:KimOhta}\\
\hline
\hline
\end{tabular}
\end{center}
\vspace{-0.5cm}
\end{table*}

\subsection{quenched spectrum in the continuum limit}
\begin{figure}[t]
\begin{center} \leavevmode
\epsfxsize=7.5cm \epsfbox{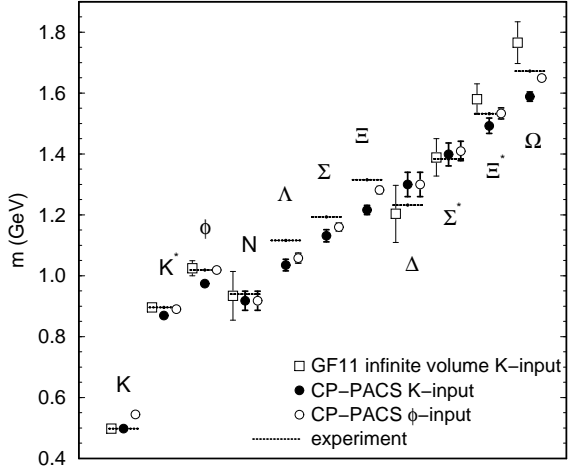}
\end{center}
\vspace{-13mm}
\caption{Quenched light hadron spectrum in the continuum limit 
reported by GF11\protect\cite{ref:GF11mass} and
CP-PACS\protect\cite{ref:CPPACS} as compared to experiment.}
\label{fig:spectrum}
\vspace{-7mm}
\end{figure}

In Fig.~\ref{fig:spectrum} we plot the result for the quenched 
light hadron spectrum reported by the CP-PACS collaboration 
as compared to the GF11 result and experiment.  
The quenched spectrum depends on the choice of hadron masses to set 
the lattice scale and light quark masses.  Results for two choices are shown 
in Fig.~\ref{fig:spectrum}, one employing $m_\pi, m_\rho$ and $m_K$ and 
the other replacing $m_K$ with $m_\phi$.
The disagreement of about 5--10\%
observed for strange hadrons between the two choices represent a 
manifestation of quenching error.

The GF11 result, albeit not covering the entire spectrum, 
showed agreement with experiment within the quoted error 
of 2\% for mesons and 4--8\% for baryons.
Comparing their result with the CP-PACS result obtained with 
the same input (filled circles), 
one finds a sizable difference for $K^*, \phi, \Xi^*$ and $\Omega$. 
In fact the CP-PACS result with significantly reduced errors 
exhibits a clear systematic deviation from experiment both for 
mesons and baryons.  

\subsection{meson spectrum}

\begin{figure}[t]
\begin{center} \leavevmode
\epsfxsize=7.5cm \epsfbox{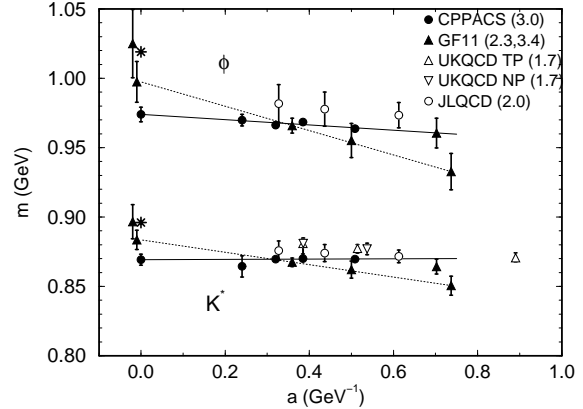}
\end{center}
\vspace{-13mm}
\caption{$m_{K^{*}}$ and $m_\phi$ with $m_K$ as input
for the Wilson (filled symbols) and clover (open symbols) 
actions. In parentheses of legends are physical lattice sizes 
in fm. Lines are continuum extrapolation adopted by GF11 and CP-PACS. 
Left-most triangles are GF11 estimates for infinite volume.
}
\label{fig:Kstar-phi}
\vspace{-7mm}
\end{figure}

The CP-PACS result in the continuum shows that the value of $m_{K^*}$ is 
3\%(6$\sigma$) smaller than experiment and $m_\phi$ by 
5\% (7$\sigma$) if $m_K$ is used as input.  Alternatively, with 
$m_\phi$ as input, they find that $m_{K^{*}}$ agrees with
experiment to 0.6\%, but $m_K$ is larger by 9\%(7$\sigma$).
This means that a small value of hyperfine splitting, previously
observed at finite lattice spacings\cite{ref:GF11mass,ref:LANLmass},
remains in the continuum limit, which is different from the conclusion 
of the GF11 collaboration after the continuum extrapolation. 

The origin of the discrepancy is clearly seen in Fig.\ref{fig:Kstar-phi}
where the continuum extrapolations of $m_{K^{*}}$ and $m_\phi$ are plotted.
The CP-PACS data (filled circles) show very small scaling violation, 
in contrast to an increase exhibited by the GF11 results. 
The continuum extrapolation of GF11 strongly
depends on the small values of results at $\beta=5.7$ obtained on a 
lattice of size $La \approx 2.3$~fm ($L=16$).  Their additional 
results for a larger lattice with $La \approx 3.4$~fm ($L=24$), also 
shown in Fig.~\ref{fig:Kstar-phi},  are higher 
by 2--3\%, and are more compatible with the CP-PACS results.  
Whether one can attribute the difference of the GF11 results 
between $L=16$ and 24 to finite-size effects 
is not clear since values of the two groups 
for smaller lattice spacings are consistent.

\begin{figure}[t]
\begin{center} \leavevmode
\epsfxsize=7.0cm \epsfbox{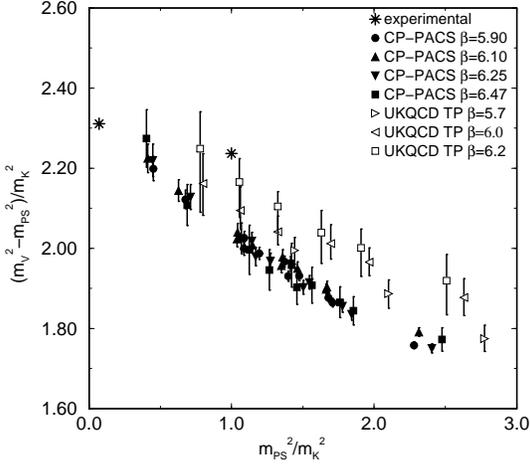}
\end{center}
\vspace{-13mm}
\caption{Meson hyperfine splitting obtained with $m_K$ as input. }
\label{fig:hyperfine}
\vspace{-7mm}
\end{figure}

In Fig.~\ref{fig:hyperfine} we plot the meson hyperfine splitting as 
a function of the pseudo-scalar meson mass squared where $m_K$ is used 
as input.  
The CP-PACS data at four values of $\beta$ (filled symbols) 
scale well and do not reproduce the experimental 
value of $K$--$K^{*}$ mass splitting.

In Figs.~\ref{fig:Kstar-phi} and \ref{fig:hyperfine}, the clover results 
have also been plotted with open symbols. 
We observe that they lie slightly above the Wilson results.
This agrees with the expectation that the clover term should increase 
the hyperfine splitting compared to that of the Wilson action. 
However, there is a problematical feature that the difference 
of results for the two actions increases toward the continuum 
limit rather than decreasing as $O(a)$. 
In fact the UKQCD collaboration\cite{ref:UKQCDlat97}
concluded this year 
that $m_{K^{*}}$ linearly extrapolated to the continuum limit is
consistent with experiment using either $m_K$ or $m_\phi$ as input. 

We should emphasize that the difference of meson masses for the two actions
is tiny(1--2\%) and no more than a 3$\sigma$ effect at finite $\beta$.
Lattice sizes of $La\simlt 2$~fm employed in the clover studies may be 
too small to avoid finite-size errors at this level of precision.
Statistical errors of the clover results,  
which are larger by a factor 2--3 compared to those of the Wilson action, 
also need to be reduced to resolve the discrepancy. 

\begin{figure}[t]
\begin{center} \leavevmode
\epsfxsize=7.5cm \epsfbox{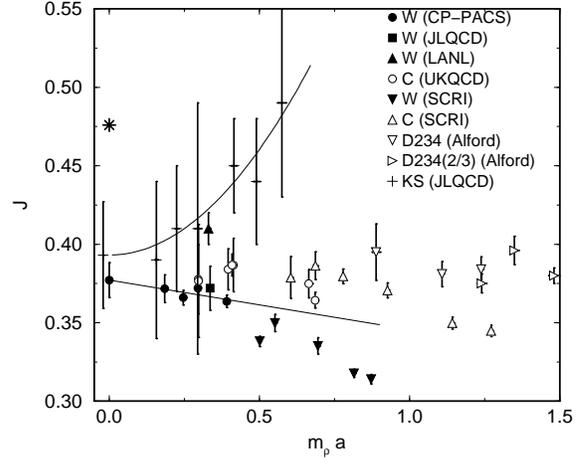}
\end{center}
\vspace{-13mm}
\caption{Results for $J$ parameter. 
Data are taken from CP-PACS\protect\cite{ref:CPPACS},
JLQCD\protect\cite{ref:JLQCD1000,ref:JLQCD-BKKS}
for the Wilson and KS actions, respectively,
LANL\protect\cite{ref:LANLmass},
SCRI\protect\cite{ref:SCRIlat96},
Alford {\it et al.}\protect\cite{ref:Alfordlat95,ref:Alfordlat96}
for the D234 and D234(2/3) actions, respectively.
Lines are fits to the CP-PACS results and the KS results.}
\label{fig:J}
\vspace{-7mm}
\end{figure}

We compile results for the $J$ parameter\cite{ref:J} in Fig.~\ref{fig:J}.  
As has been known, results for the Wilson action and its improved ones
consistently lie below the experimental value for a wide range
of lattice spacing. Results for the KS action\cite{ref:JLQCD-BKKS}
also converge to a similar value from above.
 
A small value of $J$ is equivalent to a small hyperfine splitting if the 
latter is a linear function of quark mass.
This correspondence is satisfied for the Wilson results, 
while it is apparently not for the clover case.  This represents 
another problem which needs to be understood in the quenched meson 
spectrum.

\subsection{baryon spectrum}

\begin{figure}[t]
\begin{center} \leavevmode
\vspace*{-8mm}
\epsfxsize=7.8cm \epsfbox{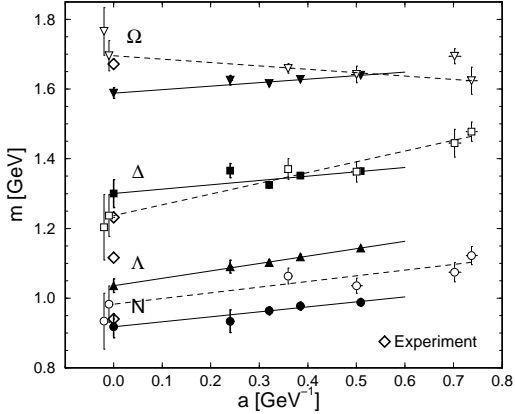}
\end{center}
\vspace{-15mm}
\caption{Continuum extrapolations of baryon masses
from CP-PACS (filled symbols) and GF11 (open symbols). } 
\label{fig:ndE}
\vspace{-6mm}
\end{figure}

In Fig.~\ref{fig:ndE} we plot the continuum extrapolation of
representative baryon masses reported by the GF11 and CP-PACS 
collaborations.  
The quenched value of nucleon mass has been a long debated issue.  
Previous high statistics results\cite{ref:APE-OLD,ref:QCDPAX,ref:LANLmass}
(see also Ref.\cite{ref:UKQCDlat97}) at $\beta\approx 5.7-6.2$ 
obtained by a chiral extrapolation 
from $m_\pi/m_\rho\simgt 0.5$ yielded a value higher than 
experiment. The GF11 results also shared this feature, and agreement 
with experiment in the continuum limit was obtained only after 
a finite-size correction.

The CP-PACS data down to $m_\pi/m_\rho\simgt 0.4$ show 
that the nucleon and $\Lambda$ masses have a negative
curvature in terms of $1/K$ toward the chiral limit.  
The bending significantly lowers the nucleon mass even at finite $\beta$
as shown in Fig.~\ref{fig:ndE}, and a linear continuum extrapolation 
leads to a value 2.3\% lower than experiment, albeit consistent 
within a 3\% statistical error.
The nucleon mass for the KS action from the MILC 
collaboration\cite{ref:MILC-Tsukuba-Gottlieb,ref:MILClat96}
is also consistent with experiment.
See Sec.~\ref{sec:chiral-N} for further discussion on the chiral 
extrapolation.

For $\Delta$ and $\Omega$ masses, the GF11 and CP-PACS results 
are reasonably consistent at similar lattice spacings. 
The continuum extrapolation is different, especially for $\Omega$, 
with the GF11 case strongly 
affected by the results at $\beta=5.7$ on an $L=16$ lattice.  

In the continuum limit, the CP-PACS results 
show a systematic deviation from experiment.  
For the octet, the non-strange nucleon mass is consistent with 
experiment, while strange baryon masses are lower 
by 5--8\% (3--5\%) with $m_K$ ($m_\phi$) as input.
However, the Gell-Mann-Okubo (GMO) relation is well
satisfied at a 1\% level.

The GMO relation is also well satisfied for the decuplet, where 
it takes the form of an equal spacing rule, 
with at most 10\% deviations.
However, the average spacing is too small by 30\% (20\%) with
$m_K$ ($m_\phi$) as input.

Baryon mass splittings were extensively studied at $\beta=6.0$ on 
a $32^3\times 64$ lattice in Ref.~\cite{ref:LANLmass}, which 
reported the validity of the GMO relations and 
the smallness of the decuplet mass splitting. 
The CP-PACS data confirm these results and extend them as the 
property of the quenched baryon spectrum in the continuum.

\subsection{quark mass for the Wilson action}
\begin{figure}[t]
\begin{center} \leavevmode
\epsfxsize=7.5cm \epsfbox{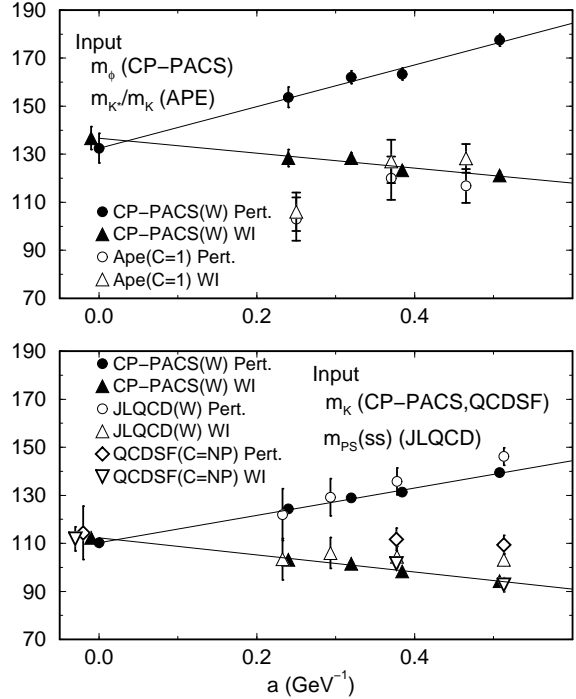}
\end{center}
\vspace{-13mm}
\caption{Comparison of strange quark masses obtained 
from the Ward identity and perturbation.
Masses are in $\overline {\rm MS}$ scheme at $\mu=2$ GeV.}
\label{fig:strange-mass}
\vspace{-8mm}
\end{figure}

The Wilson action explicitly breaks chiral symmetry at finite lattice spacing.
One of its manifestations is that quark mass $m_q^{WI}$ defined 
by the Ward identity \cite{ref:Bo,ref:Itoh86,ref:MM} 
does not agree with quark mass $m_q^P$ defined perturbatively
at finite lattice spacings\cite{ref:Itoh86,ref:LANLmass}.

This problem was examined by four groups this year. 
The CP-PACS collaboration compared the two definitions
for the Wilson action, and reported that they linearly extrapolate to 
a consistent value in the continuum limit\cite{ref:CPPACS}.  
The JLQCD collaboration employed an extended current and found indications
that scaling violation for $m_q^{WI}$ becomes smaller than that for the local 
current\cite{ref:JLQCD-Kura}.  
The QCDSF collaboration\cite{ref:QCDSFlatest} reported that
the two definitions give consistent results in the continuum limit
also for the non-perturbatively $O(a)$ improved clover action.
The Ape collaboration\cite{ref:Ape-QM} reported that $m_q^{WI}$
are compatible with $m_q^{P}$ at each $\beta$ when
renormalization factors determined non-perturbatively are used.

We summarize results for the strange quark mass in 
Fig.~\ref{fig:strange-mass}.  
The agreement of $m_q^{WI}$ with $m_q^{P}$ 
in the continuum limit supports our expectation that 
chiral symmetry of the Wilson and clover actions is recovered
in the continuum limit.
The disagreement of the values 
$m_s\approx 135$ MeV obtained with $m_\phi$ as input and 
$m_s\approx 110$ MeV found with $m_K$ as input originates 
from the small meson hyperfine splitting, and hence represents a 
quenching uncertainty.
Further results on quark masses are reviewed in Ref.~\cite{ref:Gupta}.

\section{Issues in Spectroscopic Studies}\label{sec:issues}

\subsection{finite size effects in quenched QCD}\label{sec:FS}
\begin{figure}[t]
\begin{center} \leavevmode
\epsfxsize=6.5cm \epsfbox{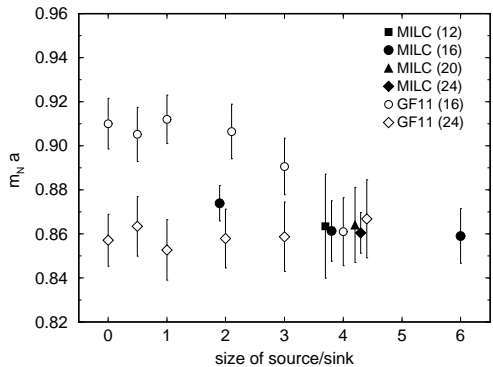}
\end{center}
\vspace{-13mm}
\caption{Nucleon mass for various source/sink and lattice sizes 
at $\beta=5.7$ and $m_\pi/m_\rho \approx 0.5$.
Data are slightly shifted in horizontal axis for clarity.
}
\label{fig:finite}
\vspace{-7mm}
\end{figure}

In quenched QCD finite-size effects of hadron masses are expected to 
be smaller than in full QCD due to Z(3) symmetry. 
For the nucleon mass with the KS action, the magnitude has been 
estimated to be less than 2\% at $m_\pi/m_\rho \approx 0.5$
for $La\simgt 2$ fm\cite{ref:Aoki-FS,ref:MILC-FS-Q}.
On the other hand, the GF11 result\cite{ref:GF11mass} for the Wilson 
action at $\beta=5.7$ showed 
a larger effect of 5\% between the sizes $L=16$ (2.3~fm) to 24 (3.4~fm).
 
The MILC collaboration carried out extensive runs 
at $\beta=5.7$ with the Wilson action for the sizes $L=12-24$, 
and we reproduce their results for 
the nucleon mass\cite{ref:MILC-Tsukuba-Gottlieb,ref:MILClat97} 
together with those of GF11 in Fig.~\ref{fig:finite}.

The GF11 result for $L=16$ significantly depends on the source/sink size, 
with the value for the size 4 consistent with those for $L=24$. 
The MILC results for $L=16$ 
do not show a source size dependence.  Their values for the sizes 
$L=12-24$ mutually agree within the statistical error of about 2\%, 
and are also consistent with the GF11 results for $L=24$.

These comparisons strongly suggest that finite-size effect at 
$La\approx 2$~fm is already 2\% or less also for the Wilson action, 
rather than 5\% estimated by GF11.  This implies that finite-size 
effects are negligible for $La \approx 3$ fm as employed by 
the CP-PACS collaboration.

\subsection{chiral extrapolation of nucleon mass}\label{sec:chiral-N}
\begin{figure}[t]
\begin{center} \leavevmode
\epsfxsize=7.5cm \epsfbox{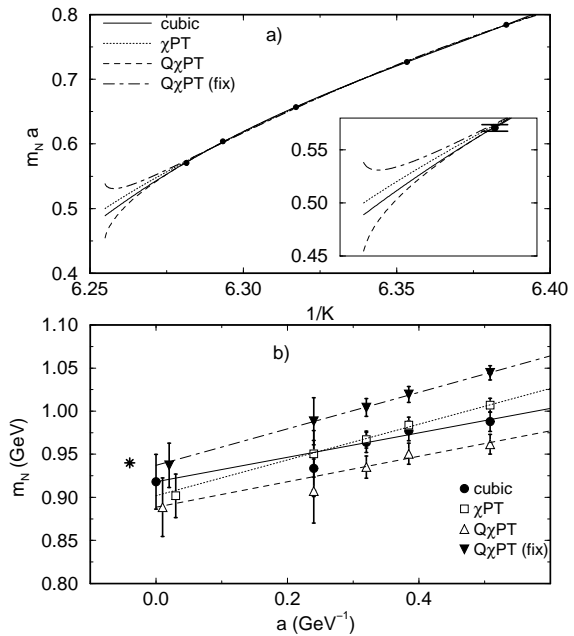}
\end{center}
\vspace{-13mm}
\caption{(a) Chiral extrapolations of the CP-PACS nucleon masses 
at $\beta=5.9$. (b) Continuum extrapolations of the nucleon masses
obtained by various chiral extrapolations.
}
\label{fig:chiral}
\vspace{-7mm}
\end{figure}

Last year the MILC 
collaboration\cite{ref:MILC-Tsukuba-Gottlieb,ref:MILClat96} 
emphasized the difficulties in reliable chiral extrapolation for the 
nucleon mass using their high precision data with the 
KS action.  The results obtained for light quarks down to 
$m_\pi/m_\rho \approx 0.3-0.4$ exhibit a negative curvature, 
and the mass in the chiral limit is sensitive to the choice of 
fitting functions.
 
The CP-PACS data for the nucleon mass for the Wilson action measured 
down to $m_\pi/m_\rho \approx 0.4$ also show a negative curvature.
They tried to fit their data using four fitting
functions; a cubic function in quark mass,
a form predicted by chiral perturbation theory ($\chi$PT)
in full QCD\cite{ref:xPT} given by  
$m_N = c_0 + c_1 m_\pi^2 + c_2 m_\pi^3$, and two forms 
in quenched QCD (Q$\chi$PT)\cite{ref:QxPT-S,ref:QxPT-BG,ref:LSc1} given by 
$m_N = c_0 + c_1 m_\pi + c_2 m_\pi^2$ and 
$m_N = c_0 - 0.53 m_\pi + c_1 m_\pi^2 + c_2 m_\pi^3$
where in the latter the coefficient of the linear term  
is fixed to a value estimated from experiment. 
As shown in Fig.~\ref{fig:chiral}(a),
the four fitting functions describe data equally well, but deviate 
significantly toward the chiral limit. 

\begin{figure}[t]
\vspace*{-3mm}
\begin{center} \leavevmode
\epsfxsize=7.0cm \epsfbox{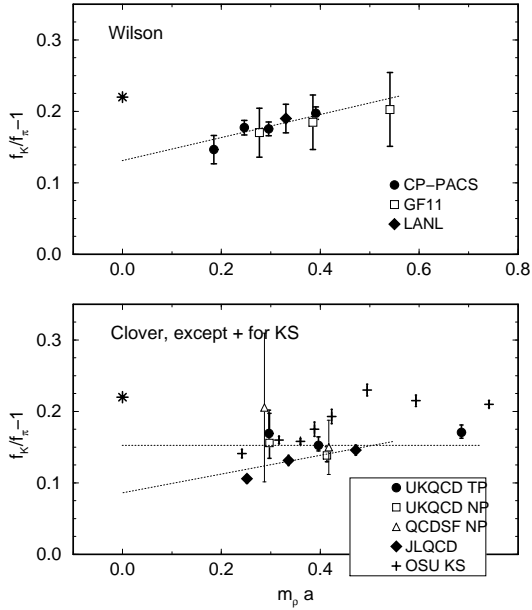}
\end{center}
\vspace{-13mm}
\caption{$f_K/f_\pi-1$. 
Data are from CP-PACS\protect\cite{ref:CPPACS},
GF11\protect\cite{ref:GF11decay},
LANL\protect\cite{ref:LANLdecay},
UKQCD\protect\cite{ref:UKQCDlat97}, 
QCDSF\protect\cite{ref:QCDSFlatest},
JLQCD\protect\cite{ref:JLQCD-fB}, and
OSU\protect\cite{ref:OSU}.}
\label{fig:decay-ratio}
\vspace{-7mm}
\end{figure}

In Fig.~\ref{fig:chiral}(b) we show how the choice of chiral extrapolations 
affects the nucleon mass in the continuum limit.  
Having precision results 
down to $m_\pi/m_\rho=0.4$ at each $\beta$ helped to constrain the 
uncertainty in the continuum limit almost within the statistical error 
of 3\%.  

A major difficulty in exploring the chiral limit in quenched QCD simulations 
is the presence of exceptional configurations.
A method has recently been proposed to avoid this 
difficulty\cite{ref:Eichten}.  It would be very interesting to see 
if the method allows to obtain reliable results near the chiral limit as 
close as $m_\pi/m_\rho \approx 0.2$, 
which would be needed to control the chiral extrapolation at a few \% 
precision level.

The APETOV collaboration\cite{ref:APETOV} studied quark mass
dependence of octet baryon masses for the non-perturbatively $O(a)$ 
improved action
for the range of $m_\pi/m_\rho = 0.98-0.56$.
They found that linearity is better 
if one includes the $O(m_qa)$ improvement term in the 
definition of quark mass.

\subsection{decay constants and quenching error}

\begin{figure}[t]
\begin{center} \leavevmode
\epsfxsize=7.0cm \epsfbox{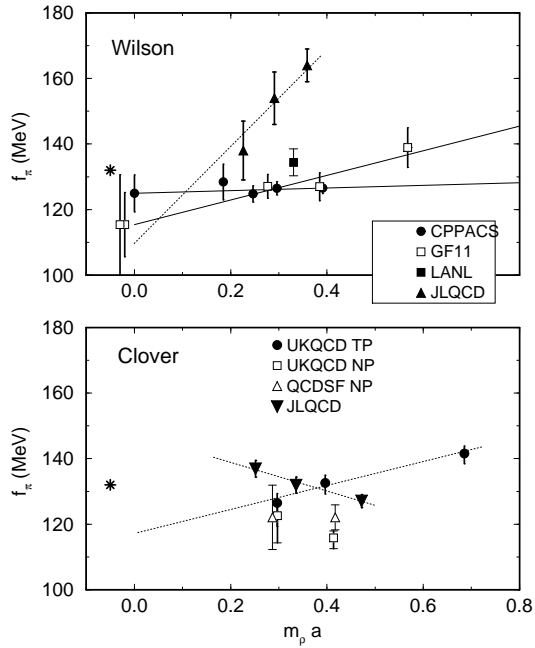}
\end{center}
\vspace{-13mm}
\caption{Results for $f_\pi$.
See the caption of Fig.~\protect\ref{fig:decay-ratio} for
references except JLQCD\protect\cite{ref:JLQCD-Kura}
for the Wilson action.}
\label{fig:decay}
\vspace{-7mm}
\end{figure}

It has been observed for the Wilson action that
$f_K/f_\pi-1$ in quenched QCD is much smaller than experiment,
which is considered to be a quenching error 
(see Ref.~\cite{ref:Sharpelat96} for a recent review).
In Fig.~\ref{fig:decay-ratio} we compile recent results for the 
ratio.  Small values in the range 0.1--0.15 are also obtained for the clover 
and KS actions.  A discrepancy of 30--40\% with experiment 
roughly agrees with estimates based on quenched chiral perturbation 
theory\cite{ref:QxPT-BG}.

In Fig.~\ref{fig:decay} we summarize the status with the determination 
of the pion decay constant.  Continuum values for the Wilson action 
reported by various groups are consistent with each other, and 
are slightly smaller than experiment, while the situation with the clover 
results is very unsatisfactory, suffering from a large discrepancy among 
groups. 

\section{Improvement of Quark Actions}\label{sec:improve}
\begin{table*}[t]
\caption{Tests of improved quark actions with improved gauge actions.
Abbreviations for gauge actions in brackets are 
TILW: tadpole-improved 
L\"uscher-Weisz\protect\cite{ref:TILW-LW,ref:LM,ref:TILW-Al},
TISY: tadpole-improved Symanzik\protect\cite{ref:Symanzik,ref:LM},
SY: Symanzik\protect\cite{ref:Symanzik}.}
\label{tab:table-I}
\begin{center}
\begin{tabular}{lrrrrrrr}
    &  $\beta_{pl}$   & size &  (fm) & \#conf. & \ \ \ $m_\pi/m_\rho$    & \#m \ 
& ref. \\
\hline
\hline
SCRI (C=NP)[TILW]   & 7.75-12 &  $8^3\times15$ & & O(1000) & & 1 &  \cite{ref:SCRIlat97} \\
\hline
Alford {\it et al.} (D234c,C)[TISY] & 1.157 & $5^3\times18$ & 2.0 & & 0.76,0.70 & 2 & 
\cite{ref:Alfordlat97}\\
Alford {\it et al.} (D234c,C)[TISY] & 1.719 & $8^3\times20$ & 2.0 & & 0.76,0.70 & 2 & 
\cite{ref:Alfordlat97}\\
\hline
DeGrand          & \multicolumn{6}{c}{Fixed point actions} & \cite{ref:DeGrand} \\
\hline
\hline
MILC  (KS,Naik)[TILW] & 7.60 & $16^3\times32$ & & 100 &0.82-0.3 & 5 & 
\cite{ref:MILClat97}\\
MILC  (KS,Naik)[TILW] & 7.75 & $16^3\times32$ & & 200 &0.76-0.33 & 5 & 
\cite{ref:MILClat97}\\
MILC  (KS,Naik)[TILW] & 7.90 & $16^3\times32$ & & 200 &0.80-0.27 & 6 & 
\cite{ref:MILClat97}\\
\hline
Bielefeld  (fat)[SY] & 4.1 & $16^3\times30$ & & 57 & $\approx 0.65$ &  & 
\cite{ref:Bielefeld}\\
\hline
\hline
\end{tabular}
\end{center}
\vspace{-7mm}
\end{table*}

\begin{figure}[t]
\begin{center} \leavevmode
\epsfxsize=7.5cm \epsfbox{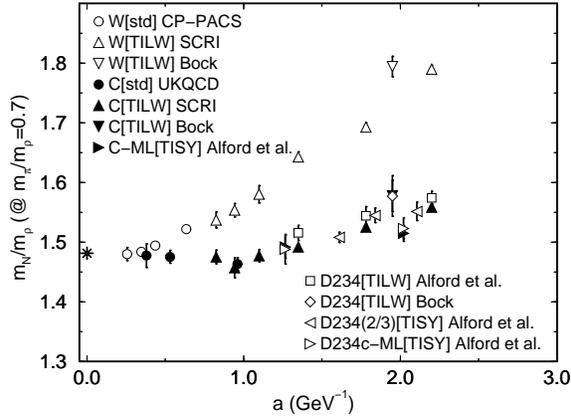}
\end{center}
\vspace{-13mm}
\caption{Comparison of $m_N/m_\rho$ at $m_\pi/m_\rho=0.7$
for various quark actions.
C-ML and D234c-ML employ mean link for the tadpole factor. 
Gauge actions are denoted in brackets.
Lattice spacings are set with the string tension 
($\protect\sqrt\sigma=427$ MeV)
except for results with TISY gauge action which use 
the charmonium spectrum.} 
\label{fig:Rho07-Q}
\vspace{-7mm}
\end{figure}

\begin{figure}[t]
\vspace*{-3mm}
\begin{center} \leavevmode
\epsfxsize=7.5cm \epsfbox{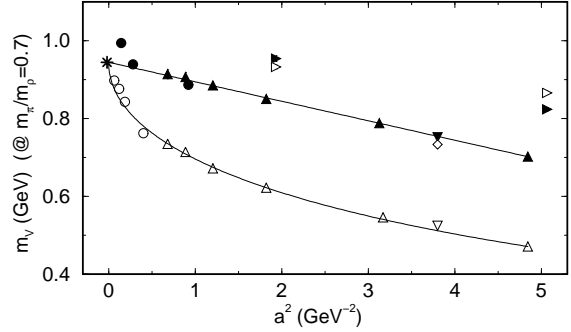}
\end{center}
\vspace{-13mm}
\caption{$m_V$ at $m_\pi/m_\rho=0.7$. Symbols are the 
same as in Fig.~\protect\ref{fig:Rho07-Q}.
Sold lines are extrapolation of SCRI 
data\protect\cite{ref:SCRIlat96}.
Lattice spacings for the C-ML and D234c-ML actions are 
recalibrated by us to those given by $\protect\sqrt\sigma$.
}
\label{fig:mv07}
\vspace{-7mm}
\end{figure}

Several groups have been testing improved quark actions with improved 
gauge actions.  In this section we discuss quenched results in this category.  
New simulations since Lattice 96 are listed in Table \ref{tab:table-I}.

\subsection{improvement of the Wilson action}

Improvement of the Wilson quark action by adding the clover term 
has been extensively investigated both
with the standard gauge 
action\cite{ref:UKQCDlat96,ref:UKQCDb57,ref:UKQCDlat97,ref:QCDSFlatest,ref:QCDSFb60,ref:QCDSFlat97-2,ref:APETOV,ref:JLQCD-fB}
and with improved gauge 
actions\cite{ref:Bock,ref:SCRIlat96,ref:Alfordlat96,ref:Alfordlat97}.
We plot in Fig.~\ref{fig:Rho07-Q} the mass ratio $m_N/m_\rho$ at 
$m_\pi/m_\rho =0.7$.  We clearly observe that the clover term 
significantly reduces scaling violation so that the ratio agrees 
with the phenomenological value\cite{ref:Ono} within 5\% 
already at $a\approx 0.4$~fm.

The D234 action\cite{ref:D234} is designed to achieve improvement  
beyond the clover action.  
Results\cite{ref:Alfordlat95,ref:Alfordlat96,ref:Alfordlat97,ref:Bock}
for a class of D234 actions, however, do not show clear improvement
for the mass ratio compared with those for the clover action.

Scaling test of hadron masses themselves at a fixed $m_\pi/m_\rho$
is useful to examine the functional dependence of scaling violation 
on the lattice spacing.
Using the tadpole-improved L\"uscher-Weisz (TILW) gauge action 
for which we expect only small scaling violation, the 
SCRI group\cite{ref:SCRIlat96} showed last year that mass results for
the tadpole-improved clover action are consistent with an 
$O(a^2)$ scaling behavior,  
while Wilson data need both $O(a)$ and $O(a^2)$ terms.
 
In Fig.\ref{fig:mv07} we reproduce their figure for the vector 
meson mass at $m_\pi/m_\rho=0.7$,
adding new results for the Wilson\cite{ref:CPPACS}(open circles) 
and clover\cite{ref:UKQCDlat97} (filled circles) actions on
the standard plaquette gauge action.
The results for the two actions lie on the respective 
extrapolation curves of the SCRI results, showing a reduction of 
scaling violation with the clover action also for the plaquette 
gauge action. 

The Cornell group\cite{ref:Alfordlat97}
tested improvement using mean value of link in the Landau 
gauge rather than plaquette for the tadpole factor(right triangles).  
They reported that the mean link is superior in reducing scaling
violation effects over plaquette.

Let us also mention that
non-perturbative determinations of
the clover coefficient with improved gauge actions 
have been attempted\cite{ref:SCRIlat97,ref:Klassen}.
Spectrum calculations are in progress.

\subsection{improvement of the KS action}

The MILC
collaboration\cite{ref:MILClat96}
studied the KS and Naik\cite{ref:Naik} three-link actions
using the TILW gauge action, and compared them with those for
the KS action on the standard gauge acton.
They found that $m_N/m_\rho$ is improved by the use of the
improved gauge action, but the Naik improvement has a relatively
small effect on the mass ratio. 
Pushing the calculation toward higher $\beta$\cite{ref:MILClat97}, 
they found little difference between the Naik and KS actions.

Another direction of improvement tested by the MILC 
collaboration\cite{ref:MILCfat} 
is the use of fat link, 
in which one replaces a link variable with a weighted sum of 
the link and staples.  
This is expected to improve flavor symmetry,
and indeed they found a substantial reduction 
in the mass difference between the Goldstone and non-Goldstone pions. 

The Bielefeld group\cite{ref:Bielefeld} studied the fat link 
improvement with the Symanzik gauge action.
They also observed improvement of flavor symmetry for this quark 
action, while $O(p^2)$ and $O(p^4)$ improved actions 
which include many link paths do not show any significant 
improvement of flavor symmetry.

\section{Toward Full QCD Spectrum}\label{sec:fullQCD}
\begin{table*}[t]
\caption{Recent spectrum runs in full QCD for $N_f$=2.
New results since Lattice 96 are marked by double asterisks and
those with increased statistics by asterisks.}
\label{tab:table-F}
\begin{center}
\begin{tabular}{lrrrrrrr}
   &       $\beta$    &  size   &  (fm) & traj.    &  $m_\pi/m_\rho$ & \#m  & ref.\\
\hline
\hline
SESAM (W)*    &  5.6   &  $16^3\times32$ &  1.4  & $200\times 25$ &  0.84-0.7 & 3 &
\cite{ref:SESAMlat96,ref:Hoeber}\\
T$\chi$L (W)* &  5.6   &  $24^3\times40$ &  2.0  & O(3000)     &  0.7,0.55 & 2 &
\cite{ref:TxLlat96,ref:Hoeber} \\
\hline
UKQCD (C=1.76)**  &  5.2  & $12^3\times24$ &     & 50 conf. & 0.85-0.75 & 4 &
\cite{ref:Talevi}\\
\hline
CP-PACS (W,C=1,TP)** &  & $(12,16)^3\times32$ & \multicolumn{4}{c}
{study of action improvement} & \cite{ref:CPPACS-F} \\
\hline
\hline
MILC (KS)  & 5.30 & $12^3\times32$ & 3.7 &1000-5000 & 0.8-0.3   & 8 &
\cite{ref:MILClat96} \\
MILC (KS)* & 5.415& $16^3\times32$ & 3.2 &1000-2000  & 0.77-0.44 & 6 &
\cite{ref:MILClat96,ref:MILClat97-F}\\
MILC (KS)* & 5.415& $12^3\times24$ & 2.4 &2000      & 0.46       & 1 &
\cite{ref:MILClat96,ref:MILClat97-F}\\
MILC (KS)  & 5.50 & $24^3\times64$ & 3.6 &1000-2000 & 0.69-0.63 & 2 &
\cite{ref:MILClat96}\\
MILC (KS)  & 5.50 & $20^3\times48$ & 3.0 &2000      & 0.56-0.48 & 2 &
\cite{ref:MILClat96}\\
MILC (KS)* & 5.60 & $24^3\times64$ & 2.6 &1500-2000 & 0.75-0.53 & 4 &
\cite{ref:MILClat96,ref:MILClat97-F}\\
\hline
Columbia(KS,$N_f$=2)*& 5.70 & $16^3\times32(40)$& 1.5 & 1400-4900 &0.70-0.57& 4 & 
\cite{ref:Columbialat96,ref:Columbialat97} \\
Columbia(KS,$N_f$=4)*& 5.40 & $16^3\times32$& 1.5 & 2700-4500&0.72-0.67& 2 &
 \cite{ref:Columbialat96,ref:Columbialat97} \\
\hline
\hline
\end{tabular}
\end{center}
\vspace*{-3mm}
\end{table*}

With progress of our understanding of the quenched spectrum, 
increasingly larger efforts are beginning to be spent in simulations of 
full QCD.  Here we summarize recent work listed in 
Table \ref{tab:table-F}.

\subsection{progress with the KS action}
\begin{figure}[t]
\begin{center} \leavevmode
\epsfxsize=7.0cm \epsfbox{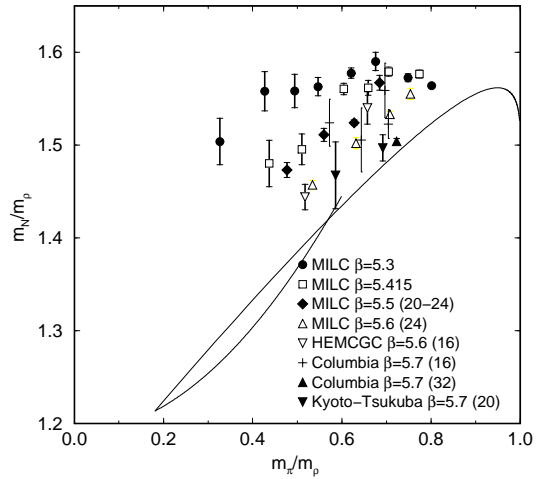}
\end{center}
\vspace{-13mm}
\caption{Edinburgh plot for $N_f$=2 KS quarks reported by 
MILC\protect\cite{ref:MILClat97-F} together with those 
from HEMCGC\protect\cite{ref:HEMCGC},
Columbia\protect\cite{ref:Columbialat97,ref:Columbia32},
and Kyoto-Tsukuba\protect\cite{ref:K-T}.
In parentheses are lattice sizes.}
\label{fig:Ed-KS-Nf2}
\vspace{-7mm}
\end{figure}

The MILC collaboration\cite{ref:MILClat96} continued their study 
of the $N_f=2$ KS spectrum for $\beta=5.3-5.6$ employing large lattices 
of a size $La \simgt 2.6$~fm. In Fig.~\ref{fig:Ed-KS-Nf2} we show their  
results in the Edinburgh plot together with those of previous 
studies\cite{ref:Columbialat97,ref:HEMCGC,ref:Columbia32,ref:K-T}. 

The ratio $m_N/m_\rho$ decreases toward weak coupling. 
Taking advantage of improved precision of their results 
as is clear from Fig.~\ref{fig:Ed-KS-Nf2}, 
the MILC collaboration attempted a continuum extrapolation of 
$m_N/m_\rho$ for a fixed value of $m_\pi/m_\rho$.  They find 
$m_N/m_\rho=1.252(37)$ at the physical point in the continuum limit. 

Also of interest is the problem of
how the KS spectrum depends on the number of dynamical 
quark flavors.
Columbia group\cite{ref:Columbialat97} showed that
the four flavor hadron spectrum is nearly parity doubled on 
a $16^3\times 32$ lattice at $\beta=5.4$.
Chiral symmetry breaking effects are smaller for four
flavors than for two or zero flavors.

\subsection{progress with the Wilson action}
Simulations of full QCD with the Wilson quark action for $N_f$=2
have been pushed forward by the SESAM\cite{ref:SESAMlat96,ref:Hoeber}
and T$\chi$L\cite{ref:TxLlat96,ref:Hoeber} collaborations.  
Simulations were initially made at $\beta=5.6$ on a $16^3$ 
spatial lattice ($La \approx 1.4$ fm) for $m_\pi/m_\rho=0.85-0.7$ (SESAM), 
which have been extended to those on a larger lattice $24^3$ 
($La \approx 2.0$ fm) and closer to the chiral limit with 
$m_\pi/m_\rho =$0.7 and 0.55 (T$\chi$L).

An important aspect of their study is a careful examination of 
various algorithmic issues of full QCD simulation, including
development and tuning of efficient Wilson marix 
inverter\cite{ref:Frommer}
and a detailed autocorrelation study. 

For the spectrum, they observed 3\% (5\%) finite-size effects
for $\rho$-meson (nucleon) at $m_\pi/m_\rho \approx 0.7$.
The magnitude is comparable to that for the KS 
action\cite{ref:K-T,ref:MILC-FS-F}.
They estimated strange hadron masses,
treating the strange quark as a valence quark
in the presence of two light dynamical quarks.
The $K-K^{*}$ mass splitting is smaller than experiment by 
15\%, contrary to the expectation that dynamical sea quark 
effects alleviate the small hyperfine splitting of quenched 
QCD.  It is possible that dynamical quarks employed is still 
too heavy to improve the splitting significantly.

SESAM and T$\chi$L also studied the static potential and several 
hadron matrix elements to explore effects of sea quarks. See 
Ref.~\cite{ref:Gusken} for a review.

\subsection{full QCD with improved actions}

Till last year there were only sporadic attempts toward full 
QCD simulations of the light hadron spectrum with improved 
actions\cite{ref:SCRIlat96-F}.
This year the CP-PACS collaboration\cite{ref:CPPACS-F} 
and the UKQCD collaboration\cite{ref:Talevi} presented 
preliminary results of a systematic attempt in this direction. 

The CP-PACS collaboration made a comparative study of improvement 
at a coarse lattice $a^{-1} \approx 0.9-1.5$ GeV employing 
the plaquette and an RG-improved action\cite{ref:Iwasaki} for gluons and
the Wilson and tadpole-improved clover action for quarks.
For one action combination, they also explored the chiral limit 
down to $m_\pi/m_\rho\approx 0.4$ with simulations on a $16^3\times 32$ 
lattice.
The UKQCD collaboration employed the plaquette action at $\beta=5.2$ and 
the clover action with a clover coefficient of 1.76.
Simulations were made for four values of sea quark masses
and the spectrum is calculated for four values of valence quark 
masses on each dynamical quark ensemble.

\begin{figure}[t]
\begin{center} \leavevmode
\epsfxsize=7.5cm \epsfbox{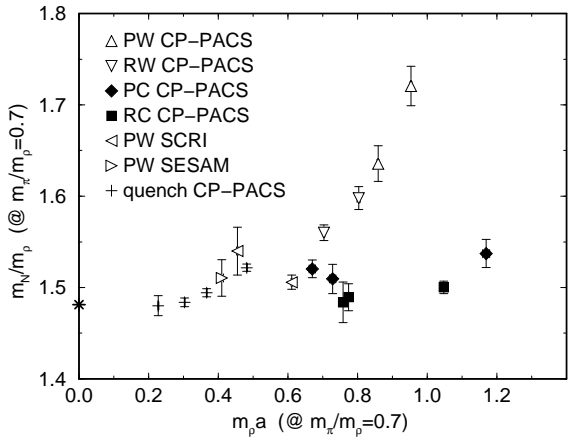}
\end{center}
\vspace{-13mm}
\caption{$m_N/m_\rho$ in full QCD with $N_f=2$ as a function of 
$m_\rho a$ both calculated at $m_\pi/m_\rho =0.7$. 
Abbreviations for gauge actions are
P: plaquette and  R: RG-improved\protect\cite{ref:Iwasaki}, and 
for quark actions W: Wilson and C: clover.
Data are taken from CP-PACS\protect\cite{ref:CPPACS-F,ref:CPPACS},
SCRI\protect\cite{ref:SCRI-W2}, and
SESAM\protect\cite{ref:Hoeber}.
}
\label{fig:rho07}
\vspace{-7mm}
\end{figure}

In Fig.\ref{fig:rho07} we compile full QCD results for $m_N/m_\rho$ 
as a function of $m_\rho a$, both calculated at $m_\pi/m_\rho =$ 0.7.
Results for the Wilson quark action have large scaling violation 
and approximately lie on a single 
curve, irrespective of the choice of gauge actions. 
In contrast the lattice spacing dependence is much 
weaker for the clover actions, again irrespective of the gauge action, 
and  the value of the ratio is close to a phenomenological estimate
even on a very coase lattice of $a^{-1} \approx 1.0$ GeV. 
These results show that a significant improvement of $m_N/m_\rho$ 
due to the clover term observed for the quenched case also 
holds in full QCD. 

Another interesting question in full QCD is to what extent 
the lattice scale obtained from the hadron spectrum agrees with 
that from the static potential.
The clover term is important also in this regard. 
A mismatch of the scale determined from $m_\rho$ in the chiral limit 
and that with the string tension observed for the Wilson action
at $a^{-1} \approx 1.0$ GeV is much reduced by the use of the clover 
action\cite{ref:CPPACS-F}.
The UKQCD collaboration reported that the scale determined from $m_{K^{*}}$ 
approximately agrees with that from $r_0$ for each value of 
dynamical quark.

For an effect of improvement of gauge actions, 
rotational symmetry of the potential is improved to a great extent
also in full QCD\cite{ref:CPPACS-F}.

The effects of improvement summarized here are parallel to those 
observed in quenched QCD, and come mainly from valence quarks rather than 
dynamical sea quarks.  
Novertheless, they are important since they show that 
realistic full QCD simulations are possible without having 
to reduce the lattice spacing below $a^{-1}\approx 2$ GeV which is  
needed with the standard action.

\section{Other Topics}\label{sec:other}
Calculation of glueball masses in quenched QCD
has reached a stage to pinpoint the mass
ranges at least for the scalar glueball. 
The GF11 collaboration\cite{ref:GF11lat97} 
reported $m_{0^{++}}=1710(63)$ MeV as the infinite volume value 
in the continuum from a reanalysis of their data\cite{ref:GF11gb}.
This value is consistent or slightly higher than the previous
results by other groups\cite{ref:UKQCDgb,ref:MPn,ref:Luo}.

The central effort of the GF11 collaboration has been a 
calculation of the mass of the $s\bar s$ scalar 
meson\cite{ref:GF11lat97,ref:GF11lat96}, for which they 
found values below $m_{s\bar s} < 1500$ MeV.
They conclude that the observed meson $f_J(1710)$ is mainly a
scalar glueball, while $f_0(1500)$ is mainly an $s\bar s$
quarkonium.  

The SESAM collaboration\cite{ref:SESAMgb}
made a glueball mass measurement with their 
full QCD runs.
No clear dynamical quark effects are seen in the glueball
masses. Instead, they observed strong finite size effects
in the scalar glueball mass, which may be an indication of
the presence of mixing between the glueball and 
the $s\bar s$ scalar meson. 

Two groups have contributions for spin exotic meson masses. 
The UKQCD collaboration\cite{ref:UKQCDlat97-ex} increased
statistics since last year. 
Calculating masses at one combination of $\beta$
and the quark mass and employing a model to estimate masses at
the strange quark,  
they obtained $m_{1^{-+}}(s\bar s) =2000(200)$ MeV.
The MILC collaboration\cite{ref:MILClat97-ex}
made simulations at $\beta$=5.85 and 6.15. 
Extrapolation to the strange quark mass was made to 
obtain $m_{1^{-+}}(s\bar s)=2170(80)$ MeV.
The two results are consistent within 10\%.

\section{Conclusions}
\label{sec:conclusions}
A number of interesting studies have been made this year, making a step
forward toward a precise determination of the light hadron spectrum.

For the quenched spectrum, a systematic deviation from experiment
has been uncovered both in the meson and baryon sectors.
Quantitative results have been accumulated with improved actions 
both for quenched and full QCD, clarifiing to what extent improving 
actions reduce scaling violations
in the light hadron spectrum.
Quenched clover simulations are moving toward high precision determination 
of physical quantities exploiting the improved scaling behavior,  and  
similar effort should be pursued with other improved actions.

And finally, attempts toward a realistic simulation in full QCD have begun.
In my opinion, there is real hope that such a calculation could be achieved 
with the current generation of computers through application
of improved actions.

\ \\
I am deeply indebted to all the colleagues who made their results available
to me before the conference.
I also would like to thank Y.~Iwasaki and A.~Ukawa for 
critical comments and suggestions on the manuscript.
This work is in part supported by the Grant-in-Aid of Ministry of
Education, Science and Culture (Nos. 08NP0101 and 09304029).


\begin{thebibliography}{99}
\bibitem{ref:reviews}
For recent reviews, see S.~Gottlieb, \lats 155;
D.~K.~Sinclair, \latf 112.
\bibitem{ref:MILC-Tsukuba-Gottlieb}
MILC Collaboration, C.~Bernard {\it et al.}, hep-lat/9707014.
\bibitem{ref:MILClat97}
MILC Collaboration, talk by S.~Gottlieb, these proceedings.
\bibitem{ref:CPPACS}
CP-PACS Collaboration, 
talk by K.~Kanaya, these proceedings.
\bibitem{ref:UKQCDlat96}
UKQCD Collaboration, R.~Kenway {\it et al.}, \lats 206.
\bibitem{ref:UKQCDb57}
UKQCD Collaboration, H~.~Shanahan {\it et al.}, 
Phys. Rev. D55 (1997) 1548.
\bibitem{ref:UKQCDlat97}
UKQCD Collaboration, 
talk by P.~Rowland, these proceedings.
\bibitem{ref:QCDSFlatest}
QCDSF Collaboration, M.~G\"ockeler {\it et al.}, hep-lat/9707021;
talk by P.~Stephenson, these proceedings.
\bibitem{ref:QCDSFb60}
QCDSF Collaboration, M.~G\"ockeler {\it et al.}, Phys. Lett. B391 (1997) 388.
\bibitem{ref:QCDSFlat97-2}
QCDSF Collaboration, poster by D.~Pleiter, these proceedings. 
\bibitem{ref:APETOV}
APETOV Collaboration, talk by T.~Mendes, these proceedings.  
\bibitem{ref:JLQCD-fB}
JLQCD Collaboration, talk by S.~Hashimoto, these proceedings.
\bibitem{ref:KimOhta}
S.~Kim and S.~Ohta, \lats 199; poster by S.~Kim, these proceedings.
\bibitem{ref:GF11mass}
F.~Butler, {\it et al.},Nucl. Phys. B430 (1994) 179.
\bibitem{ref:SW}
B.~Sheikholeslami and R.~Wohlert, Nucl. Phys. B259 (1985) 572.
\bibitem{ref:LM}
G.~P.~Lepage and P.~B.~Mackenzie, Phys. Rev. D48 (1993) 2250.
\bibitem{ref:NPI}
M.L\"uscher {\it et al.}, Nucl.Phys.B491 (1997) 323. 
\bibitem{ref:Wittig}
H.~Wittig, review in these proceedings.
\bibitem{ref:MILClat96}
MILC Collaboration, C.~Bernard {\it et al.}, \lats 212.
\bibitem{ref:LANLmass}
T.~Bhattacharya, R.~Gupta, G.~Kilcup and S.~Sharpe,
Phys. Rev. D53 (1996) 6486.
\bibitem{ref:J}
UKQCD Collaboration, P.~Lacock and C.~Michael,
Phys.~Rev. D52 (1995) 5213.
\bibitem{ref:JLQCD1000}
JLQCD Collaboration, S.~Aoki {\it et al.},
Nucl. Phys. B (Proc. Suppl.) 47 (1996) 354.
\bibitem{ref:JLQCD-BKKS}
JLQCD Collaboration, S.~Aoki {\it et al.},
\lats 341; talk by S.~Aoki, these proceedings.
\bibitem{ref:SCRIlat96}
S.~Collins {\it et al.}, \lats 877; hep-lat/9710021.
\bibitem{ref:Alfordlat95}
M.~Alford, K.~Klassen and P.~Lepage, \latf 370.
\bibitem{ref:Alfordlat96}
M.~Alford, T.~Klassen and P.~Lepage, \lats 861.
\bibitem{ref:APE-OLD}
P.~Bacilieri {\it et al.}, Nucl.Phys.B317 (1989) 509; 
S.~Cabasino {\it et al.}, Phys.Lett.B258 (1991) 195.
\bibitem{ref:QCDPAX}
QCDPAX Collaboration, Y.~Iwasaki {\it et al.},
Phys. Rev. D53 (1996) 6443.
\bibitem{ref:Bo}
M.Bochicchio {\it et al.}, Nucl.Phys.B262 (1985) 331.
\bibitem{ref:Itoh86}
S.~Itoh {\it et al.}, Nucl.Phys. B274 (1986) 33.
\bibitem{ref:MM}
L.~Maiani and G.~Martinelli, Phys.~Lett. B178 (1986) 265.
\bibitem{ref:JLQCD-Kura}
JLQCD Collaboration, poster by \\Y.~Kuramashi, these proceedings.
\bibitem{ref:Ape-QM}
Ape Collaboration, talk by L.~Giusti, these proceedings.
\bibitem{ref:Gupta}
R.~Gupta, review in these proceedings. 
\bibitem{ref:Aoki-FS}
S.~Aoki {\it et al.}, Phys. Rev. D50 (1994) 486.
\bibitem{ref:MILC-FS-Q}
S.~Gottlieb, Nucl. Phys. B (Proc.Suppl.) 42 (1995) 346.
\bibitem{ref:xPT}
J.~Gasser and H.~Leutwyler, Phys. Rep. C87 (1982) 77;
Nucl. Phys. B250 (1985) 465. 
\bibitem{ref:QxPT-S}
S.~R.~Sharpe, Phys.~Rev. D41 (1990) 3233;
{\it ibid.} D46 (1992) 3146. 
\bibitem{ref:QxPT-BG}
C.~Bernard and M.~Golterman, Phys. Rev. D46 (1992) 853; 
Nucl. Phys. B (Proc. Suppl.) 26 (1992) 360; {\it ibid.} 30 (1993) 217.
\bibitem{ref:LSc1}
J.~N.~Labrenz and S.~R.~Sharpe,
Nucl. Phys. B (Proc. Suppl.) 34 (1994) 335; 
Phys. Rev. D54 (1996) 4595.
\bibitem{ref:Eichten}
E.~Eichten, review in these proceedings, and references
therin. 
\bibitem{ref:Sharpelat96}
For a recent review, see 
S.~R.~Sharpe, \lats 181.
\bibitem{ref:GF11decay}
F.~Butler, {\it et al.}, Nucl. Phys. B421 (1994) 217.
\bibitem{ref:LANLdecay}
T.~Bhattacharya and R.~Gupta, Phys. Rev. D54 (1996) 1155.
\bibitem{ref:OSU}
G.~Kilcup and D.~Pekurovsky, private communications.
\bibitem{ref:TILW-LW}
M.~L\"uscher and P.~Weisz, Comm. Math. Phys. 97 (1985) 59;
Phys. Lett. B158 (1985) 250.
\bibitem{ref:TILW-Al}
M.~Alford {\it et al.}, Phys. Lett. B361 (1995) 87.
\bibitem{ref:Symanzik}
K.Symanzik, Nucl.Phys.B226 (1983) 187,205.
\bibitem{ref:SCRIlat97}
R.G.Edwards, U.M.Heller and T.R.Klassen, these proceedings. 
\bibitem{ref:Alfordlat97}
M.~Alford, T.~Klassen and P.~Lepage, these proceedings.
\bibitem{ref:DeGrand}
T.~DeGrand, these proceedings.
\bibitem{ref:Bielefeld}
A.~Peikert {\it et al.}, these proceedings.
\bibitem{ref:Bock}
W.~Bock, \lats 870.
\bibitem{ref:Ono}
S.~Ono, Phys. Rev. D17 (1978) 888. 
\bibitem{ref:D234}
M.~Alford, T.~Klassen and P.~Lepage, Nucl. Phys. B496 (1997) 377,
and references therein.
\bibitem{ref:Klassen}
T.~Klassen, these proceedings.
\bibitem{ref:Naik}
S.~Naik, Nucl. Phys. B316 (1989) 238. 
\bibitem{ref:MILCfat}
MILC Collaboration, T.~Blum {\it et al.}, Phys. Rev. D55 (1997) 1133.

\bibitem{ref:SESAMlat96}
SESAM Collaboration, U.~Gl\"assner {\it et al.},
\lats 219.
\bibitem{ref:TxLlat96}
T$\chi$L Collaboration, L.~Conti {\it et al.},
\lats 222.
\bibitem{ref:Hoeber}
SESAM and T$\chi$L Collaborations, talk by H.~Hoeber, these proceedings.
\bibitem{ref:Talevi}
UKQCD Collaboration, talk by M.~Talevi, these proceedings.
\bibitem{ref:CPPACS-F}
CP-PACS Collaboration,
talks by \\R.Burkhalter and T.Kaneko,\\these proceedings. 
\bibitem{ref:MILClat97-F}
MILC Collaboration, poster by R.~Sugar, these proceedings.
\bibitem{ref:Columbialat96}
D.Chen and R.D.Mawhinney, \lats 216.
\bibitem{ref:Columbialat97}
R.~D.~Mawhinney, hep-lat/9705031 ; these proceedings.
\bibitem{ref:HEMCGC}
K.M.Bitar {\it et al.}, Phys.Rev.D49 (1994) 6026.
\bibitem{ref:Columbia32}
W.~Schaffer, Nucl. Phys. B(Proc.Suppl.) 30 (1993) 405. 
\bibitem{ref:K-T}
M.Fukugita {\it et al.}, Phys.Rev.D47 (1993) 4739. 
\bibitem{ref:Frommer}
A.~Frommer {\it et al.}, Int. J. Mod. Phys. C5 (1994) 1073.
\bibitem{ref:MILC-FS-F}
C.~Bernard {\it et al.}, Phys.Rev.D48 (1993) 4419; 
Nucl. Phys. B (Proc.Suppl.) 34 (1994) 366.
\bibitem{ref:Gusken}
S.~G\"usken, review in these proceedings.
\bibitem{ref:SCRIlat96-F}
S.~Collins {\it et al.}, \lats 880.
\bibitem{ref:Iwasaki}
Y.~Iwasaki, Nucl. Phys. B258 (1985) 141;
Univ. of Tsukuba report UTHEP-118 (1983).
\bibitem{ref:SCRI-W2}
K.~M.~Bitar {\it et al.}, Phys.Rev.D54 (1996) 3546.
\bibitem{ref:GF11lat97}
W.Lee and D.Weingarten, these proceedings.
\bibitem{ref:GF11gb}
H.~Chen {\it et al.}, Nucl. Phys. B (Proc. Suppl.) 34 (1994) 357.
\bibitem{ref:UKQCDgb}
UKQCD Collaboration, G.Bali {\it et.al.}, Phys. Lett. B309 (1993) 378. 
\bibitem{ref:MPn}
C.~J.~Morningstar and M.~Peardon, Phys.Rev. D56 (1997) 4043.
\bibitem{ref:Luo}
X.~Q.~Luo and Q. Chen, Mod. Phys. Lett. A11 (1996) 2435;
X.~Q.~Luo {\it et al.}, \lats 243.
\bibitem{ref:GF11lat96}
D.~Weingarten, \lats 232; W.~Lee and D.~Weingarten, \lats 236.
\bibitem{ref:SESAMgb}
SESAM Collaboration, talk by G.~Bali, these proceedings.
\bibitem{ref:UKQCDlat97-ex}
UKQCD Collaboration, P.~Lacock {\it et al.},
Phys. Lett. B401 (1997) 308; talk by\\ P.~Lacock, these proceedings. 
\bibitem{ref:MILClat97-ex}
MILC Collaboration, C.~Bernard {\it et al.}, \lats 228; hep-lat/9707008; 
poster by D.~Toussaint, these proceedings. 

\end{thebibliography}
\end{document}